\documentclass[
aip,
amsmath,amssymb,
reprint,
]{revtex4-1}

\usepackage[colorlinks, urlcolor=blue, linkcolor=blue, citecolor=blue]{hyperref}
\usepackage{graphicx}
\usepackage{dcolumn}
\usepackage{bm}
\usepackage[utf8]{inputenc}
\usepackage[T1]{fontenc}
\usepackage{mathptmx}
\usepackage{amsmath,amsfonts,amssymb}
\usepackage{float}
\usepackage[squaren]{SIunits}
\usepackage{textcomp}
\usepackage{romannum}
\usepackage[utf8]{inputenc}
\usepackage[T1]{fontenc}
\usepackage{xcolor}
\usepackage{color}
\usepackage{multirow}
\usepackage[squaren]{SIunits}
\graphicspath{{figures/}}
\usepackage{hyperref}

\begin{document}

\typeout{COLUMNWIDTH=\the\columnwidth}

\title{High-transformer-ratio plasma-based afterburner for extending X-ray free-electron laser photon energies to $\gtrsim$50 keV}

\author{Songtang Han}
\affiliation{
State Key Laboratory of Nuclear Physics and Technology, and Key Laboratory of HEDP of the Ministry of Education, CLAPA, Peking University, Beijing 100871, China
}

\author{Letian Liu}
\affiliation{
State Key Laboratory of Nuclear Physics and Technology, and Key Laboratory of HEDP of the Ministry of Education, CLAPA, Peking University, Beijing 100871, China
}

\author{Zhiyan Yang}
\affiliation{
State Key Laboratory of Nuclear Physics and Technology, and Key Laboratory of HEDP of the Ministry of Education, CLAPA, Peking University, Beijing 100871, China
}

\author{Zhuo Pan}
\affiliation{
State Key Laboratory of Nuclear Physics and Technology, and Key Laboratory of HEDP of the Ministry of Education, CLAPA, Peking University, Beijing 100871, China
}

\author{Xueqing Yan}
\affiliation{
State Key Laboratory of Nuclear Physics and Technology, and Key Laboratory of HEDP of the Ministry of Education, CLAPA, Peking University, Beijing 100871, China
}
\affiliation{
Beijing Laser Acceleration Innovation Center, Huairou, Beijing 101400, China
}
\affiliation{
Institute of Guangdong Laser Plasma Technology, Baiyun, Guangzhou 510540, China
}

\author{Xinlu Xu}
\email[Author to whom correspondence should be addressed: ]{xuxinlu@pku.edu.cn}
\affiliation{
State Key Laboratory of Nuclear Physics and Technology, and Key Laboratory of HEDP of the Ministry of Education, CLAPA, Peking University, Beijing 100871, China
}
\affiliation{
Beijing Laser Acceleration Innovation Center, Huairou, Beijing 101400, China
}

\date{\today}

\begin{abstract}
Bright coherent X-rays in the $\gtrsim$50 keV range are highly desirable for applications in materials science, high-energy-density physics, and nuclear physics, but are difficult to obtain from present X-ray free-electron laser (XFEL) facilities. Plasma-based afterburners offer a compact route to increasing the electron-beam energy, yet high transformer ratio (TR) operation is vulnerable to hosing instability, which can severely degrade the witness beam quality. We propose using a transversely elliptical driver beam to excite an elliptical nonlinear plasma wake with an axially varying focusing gradient, thereby suppressing the witness beam hosing through betatron decoherence. Three-dimensional particle-in-cell simulations show that, for a 3 kA driver and a 2.35 kA witness beam with an rms length of 4 $\micro\meter$ and a beam separation of 94 $\micro\meter$ ($\text{TR}\approx 3$), this scheme can accelerate an 8 GeV witness beam to $\sim30$ GeV while nearly half of the electrons meet the XFEL requirements. Start-to-end simulations show that the resulting beam can drive $\gtrsim 50~\mathrm{keV}$ XFEL radiation with peak powers of $\sim 10$ GW and durations of $\sim 10$ fs, even when quantum diffusion is included, offering a compact route to extend existing XFEL facilities to higher photon energies.
\end{abstract}

\maketitle

\section{Introduction}
X-ray free-electron lasers (XFELs) can deliver X-rays with unprecedented brightness \cite{pellegrini2016physics,FeatureFEL2020}, ultrashort pulse duration, and exceptional transverse coherence, enabling applications across many disciplines, such as nonlinear X-ray physics \cite{young2010femtosecond, fuchs2015anomalous,PhysRevLett.111.233902}, ultrafast science \cite{yong2020observation,loh2020observation,pande2016femtosecond}, and coherent diffractive imaging \cite{chapman2006femtosecond,chapman2011femtosecond,PhysRevLett.114.098102,boutet2012high}. Limited by facility footprint and construction cost, most XFELs employ high‑quality electron beams with energies of $\sim$10 GeV to generate X‑ray photons of $\sim1$ to $\sim20$ keV in magnetic undulators with a length of an order of 100 m \cite{emma2010first, schneidmiller2017baseline,FeatureFEL2020, yabashi2015overview, milne2017swissfel, stohr2011linac, ko2017construction, zhu2017sclf}. Extending the photon energy to the several 10s to 100 keV range is highly desirable \cite{tan2025arXiv} for advancing applications in fields such as deep-penetration materials science \cite{lienert2011high,bernier2020high,mimura2022high}, physics under extreme conditions \cite{zastrau2021high}, and nuclear quantum optics \cite{adams2013x}. For example, high-energy diffraction microscopy (HEDM) utilizes 50$\sim$100 keV X-rays to resolve grain-level three-dimensional structure and dynamics within bulk polycrystalline materials, which is beyond the reach of current XFELs \cite{lienert2011high,bernier2020high}.

Shortening the undulator period $\lambda_\mathrm{u}$ or using a smaller undulator deflection parameter $K$ can increase the XFEL photon energy, but at the expense of a longer saturation length and reduced saturation power \cite{pellegrini2016physics, HuangRevofFEL2007}. In contrast,  increasing the electron beam energy raises the photon energy without decreasing the saturation power, making it the most effective approach. Plasma-based afterburners (PBAs) \cite{AfterburnerSLAC2002,blumenfeld2007energy,litos2014high, DArcy2022Nature} with $\sim 10~\giga\volt\per\meter$ acceleration gradient have been proposed as a means of boosting the electron beam energy at XFEL facilities by avoiding substantially increasing the facility footprint and cost \cite{Schroder_2025,liu2025improvement}. In a PBA, one electron beam from the radio-frequency (RF) accelerators is used as a driver to excite a nonlinear plasma wake \cite{PhysRevLett.43.267,chen1985acceleration,RevModPhys.81.1229} and accelerate another RF accelerator beam - the witness beam. Although they have demonstrated 10-GeV-class energy gains in meter-long plasmas \cite{litos2014high,blumenfeld2007energy, Litos_2016}, it is challenging to preserve the beam quality to meet the stringent requirements posed by XFELs, i.e., $\epsilon_\mathrm{n} \lesssim \gamma_\mathrm{b}\frac{\lambda_\mathrm{r}}{4\pi}$ and $\frac{\Delta {\gamma_\mathrm{b}}}{\gamma_\mathrm{b}} < \rho$, where $\lambda_\mathrm{r}$ is the radiation wavelength, $\rho$ is the dimensionless Pierce parameter characterizing FEL gain, $\epsilon_\mathrm{n}, \Delta\gamma_\mathrm{b} mc^2, \gamma_\mathrm{b} mc^2$ are the normalized emittance, the slice energy spread and the average energy of the electron beam, respectively, $m$ is the electron mass and $c$ is the speed of light in vacuum \cite{pellegrini2016physics,HuangRevofFEL2007}. 

A recent work \cite{liu2025improvement} studied the possible reasons of the beam quality degradation in a PBA, including axial chirp of the acceleration gradient \cite{PhysRevA.44.R6189,Kostyukov2004PoP,PhysRevLett.96.165002} and ion motion \cite{PhysRevLett.95.195002,PhysRevLett.104.155001,IonMotion2017}, and showed that more than half of the witness beam can satisfy the subsequent XFEL requirements when utilizing the beam loading \cite{PhysRevLett.101.145002}. The transformer ratio (TR) \cite{ruth1984plasma,Bane1985IEEE, PhysRevLett.56.1252, PhysRevLett.58.555} which is defined as the ratio of the maximum accelerating field experienced by the witness beam to the maximum decelerating field acting on the driver reaches $\sim 1$. To further improve the witness beam energy gain and the corresponding XFEL photon energy, a PBA operated with a large TR is needed which is susceptible to hosing instability (HI) \cite{Whittum1991, Huang2007} seeded by misalignment between the driver and witness beams and/or the asymmetry of the beams \cite{PhysRevLett.121.064801, PhysRevLett.124.044802}. HI would wiggle the beams, amplify the transverse position and angle jitters, and degrade the beam quality rapidly, and it is thus necessary to suppress it. Several schemes have been proposed to suppress the HI through a decoherence of the electron betatron oscillations, e.g., applying a large energy chirp \cite{PhysRevLett.118.174801, 2018IEEEmitigation}, increasing the slice energy spread \cite{PhysRevLett.118.174801, PhysRevAccelBeams.22.031302,Finnerud2025instability}, utilizing the axial varying focusing gradient in the quasilinear regime \cite{PhysRevLett.119.244801} and a substantial ion motion \cite{PhysRevLett.121.264802,Hildebrand2025realignment}. However, these schemes all degrade the beam quality and are therefore only applicable to suppressing the HI of the driver beam. 

In this work, a transversely elliptical driver beam is employed to excite an elliptical nonlinear plasma wake with an axially varying focusing gradient, thereby mitigating the HI of the witness beam and maintaining its quality. We combine theoretical analysis with three-dimensional (3D) particle-in-cell (PIC) simulations to investigate the performance of the TR of the PBA and the usable fraction of the witness beam under different parameters of the driver and the witness beams. Simulations indicate that, in the axisymmetric wake generated by a round driver at $\mathrm{TR}\approx3$, the HI seeded by the beam misalignment can cause significant particle loss and catastrophic quality degradation. In contrast, driver beams optimized in spot size and ellipticity can suppress HI almost completely with limited growth of the slice energy spread. Under these conditions the initial 8 GeV witness beam can be accelerated to $\sim 30$ GeV while retaining its initial quality, enabling the production of $\gtrsim$50 keV x-rays with 10s GW peak power, even after accounting for quantum diffusion. The proposed PBA scheme can be integrated into contemporary XFEL facilities, substantially extending their capability for high-photon-energy x-ray generation and benefiting hard-x-ray users in nuclear physics, high-energy-density physics, and materials science.

\section{High-quality PBA with a high TR}
\begin{figure}[!t]
    \centering
    \includegraphics[width=1\linewidth]{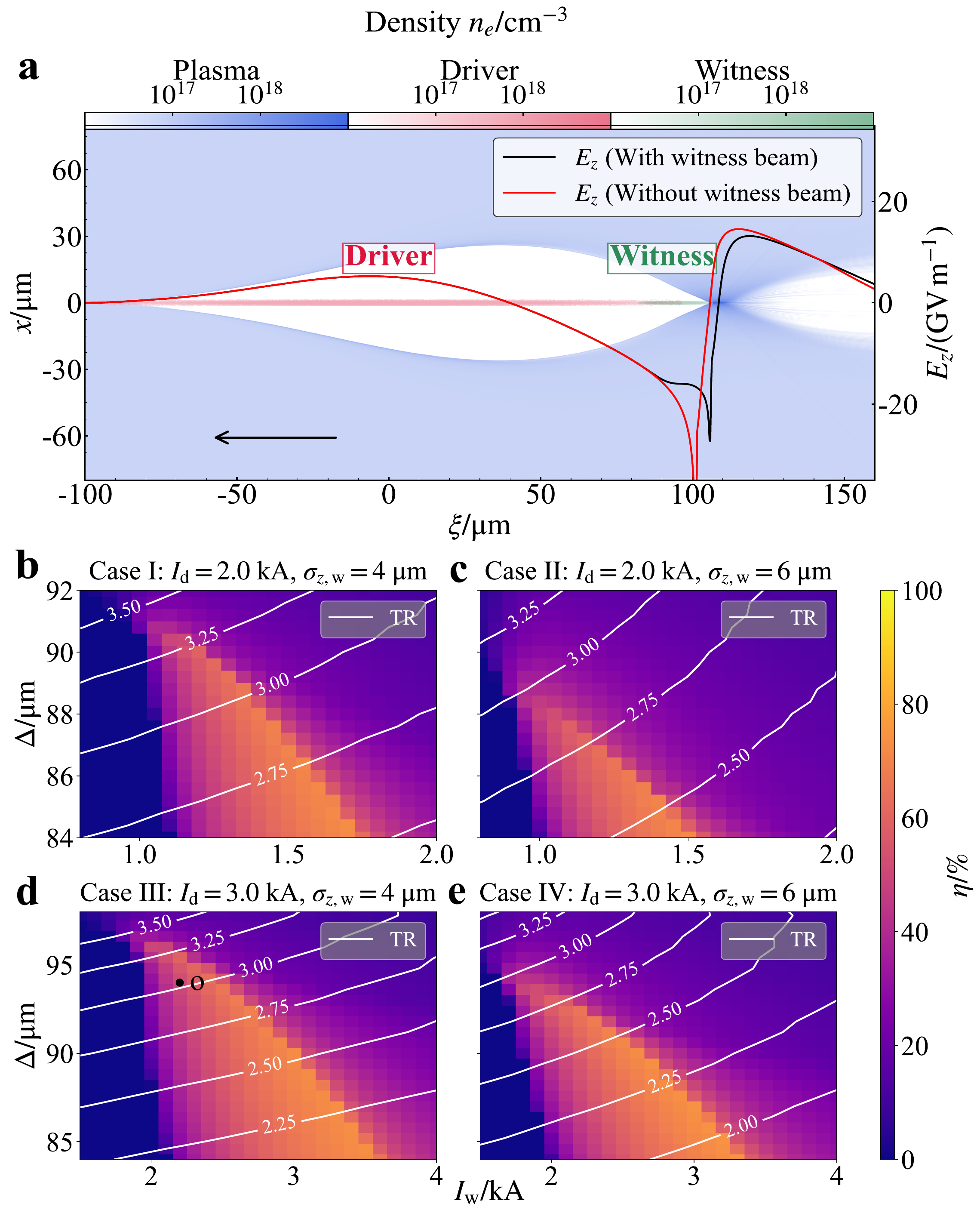}
    \caption{Schematic and parameter scan of the two-bunch PBA. (a) Plasma electron density distribution together with the driver and witness beams. The plasma electron density is shown in blue, while the driver- and witness-beam density distributions are shown in red and green, respectively. The black curve shows the on-axis accelerating field $E_z$ with the witness beam included, and the red curve shows the corresponding unloaded wake. (b)--(e) Dependence of the transformer ratio and usable fraction $\eta$ on the witness beam peak current $I_\mathrm{w}$ and beam separation $\Delta$ for four cases. White contours denote the transformer ratio, and color maps denote $\eta$. For panels (b) and (c), the driver current is fixed at $I_\mathrm{d}=2.0~\mathrm{kA}$, with scan step sizes of $0.05~\mathrm{kA}$ in $I_\mathrm{w}$ and $0.4~\micro\meter$ in $\Delta$. For panels (d) and (e), the driver current is fixed at $I_\mathrm{d}=3.0~\mathrm{kA}$, with scan step sizes of $0.1~\mathrm{kA}$ in $I_\mathrm{w}$ and $0.6~\micro\meter$ in $\Delta$.}  
    \label{fig:wakefield_and_parameter_scan}
\end{figure}

We consider contemporary XFEL facilities and set the energy of the electron beams from the RF accelerators to 8 GeV without loss of generality \cite{emma2010first, schneidmiller2017baseline, FeatureFEL2020, yabashi2015overview, milne2017swissfel, stohr2011linac, ko2017construction, zhu2017sclf}. As shown in Fig. \ref{fig:wakefield_and_parameter_scan}(a), two temporally separated electron beams propagate into a uniform plasma with a density of $n_0=7 \times 10^{16}~\mathrm{cm^{-3}}$ where the driver transfers its energy to the witness beam through the nonlinear plasma wake. In the linear regime of plasma wakefield acceleration, the TR is limited to 2 for longitudinally symmetric drivers \cite{ruth1984plasma}. Asymmetric drivers with slowly rising edge and rapidly falling edge can flatten the decelerating field and achieve a high TR which exceeds 2 \cite{Bane1985IEEE,PhysRevLett.56.1252}. In the 3D nonlinear blowout regime \cite{PhysRevLett.96.165002,PhysRevA.44.R6189, Pukhov2002}, a longitudinally symmetric beam driver with relatively long duration can also achieve a large TR since the profile of the acceleration field is governed by the motion of the boundary of the nonlinear wake rather than the falling edge of the driver \cite{lu2009high, Su2023PoP,PhysRevLett.121.064801, PhysRevLett.124.044802}. 

To avoid the additional complexity associated with beam shaping, we consider tri-Gaussian distribution for both driver and witness beams in the subsequent discussions. We choose a driver with a root-mean-square (rms) length of $\sigma_{z,\mathrm{d}}=32~\mu\mathrm{m}$ ($k_\mathrm{p}\sigma_{z,\mathrm{d}}=1.6$) and consider four representative 
configurations of driver peak current and witness bunch length: Case~\Romannum{1} ($I_\mathrm{d}=2~\mathrm{kA}$, $\sigma_{z,\mathrm{w}}=4~\micro\meter$), Case~\Romannum{2} ($I_\mathrm{d}=2~\mathrm{kA}$, $\sigma_{z,\mathrm{w}}=6~\micro\meter$), Case~\Romannum{3} ($I_\mathrm{d}=3~\mathrm{kA}$, $\sigma_{z,\mathrm{w}}=4~\micro\meter$), and Case~\Romannum{4}($I_\mathrm{d}=3~\mathrm{kA}$, $\sigma_{z,\mathrm{w}}=6~\micro\meter$), where $k_\mathrm{p} = \frac{\omega_\mathrm{p}}{c}$, $\omega_\mathrm{p} = \sqrt{\frac{n_0 e^{2}}{m \epsilon_0}}$ is the plasma frequency, $e$ is the electron charge and $\epsilon_0$ is the vacuum permittivity. For each case, we scan the peak current of the witness ($I_\mathrm{w}$) and the beam separation ($\Delta$) to investigate the TR and the preservation of witness beam quality. The beams are assumed to be transversely matched to the ion column force of the nonlinear plasma wake, i.e., their Twiss parameters satisfy $\beta=\sqrt{2\gamma_\mathrm{b}} k_\mathrm{p}^{-1} \approx 3.6$~mm and $\alpha=0$ \cite{PhysRevLett.112.035003}. Achieving this millimeter-scale $\beta$ using magnetic focusing elements is challenging; however, this can be addressed by employing a properly tailored plasma density upramp to further focus the beam to the required $\beta$ \cite{xu2016matching, PhysRevAccelBeams.23.011302}. The initial emittance of the witness beam is set to 0.4 $\micro\meter$, representative of state-of-the-art beam quality, while a larger value of 1.2 $\micro\meter$ is assigned to the driver. Their matched spot sizes are $\sigma_{r,\mathrm{d}}=0.52~\micro\meter$ and $\sigma_{r,\mathrm{w}}=0.30~\micro\meter$. The plasma length is chosen such that the driver energy is almost fully depleted. 
    
We use 3D quasi-static PIC code QuickPIC to model the PBA stage \cite{QuickPIC2006,an2013improved}. The simulation setup can be found in Appendix A. The tightly focused driver beam with a peak density of $n_\mathrm{d}\approx2.5\times 10^{19}~\centi\meter^{-3} \gg n_0$ (for $I_\mathrm{d}=2~\mathrm{kA}$) excites a fully blowout plasma wake where an ion column void of plasma electrons is present [Fig. \ref{fig:wakefield_and_parameter_scan}(a)]. The transverse wakefield in the ion column is linear as $W_r=-\frac{1}{2}\frac{m\omega_\mathrm{p}^2}{e}r$, and the acceleration gradient $E_z$ is uniform in the transverse plane but varies longitudinally \cite{Kostyukov2004PoP,PhysRevLett.96.165002}. The linear focusing force with a constant gradient ($\frac{W_r}{r}$) can preserve the beam emittance \cite{PhysRevLett.112.035003,Lindstrom2024NC}. However, the axial variation of $E_z$ increases the energy spread within a finite-thickness longitudinal slice of the witness beam. In FELs, electrons within one cooperation length, $l_\mathrm{c} \approx 20 \frac{\lambda_\mathrm{r}}{4\pi\sqrt{3}\rho}$, emit radiation cooperatively \cite{pellegrini2016physics}. Consequently, the relative energy spread of a slice with thickness $l_\mathrm{c}$ must satisfy $\frac{\sigma_{\gamma_\mathrm{b}}}{\gamma_\mathrm{b}}<\rho$ to maintain FEL gain, where $\sigma$ denotes the rms value. By assuming a non-evolving $E_z$ profile (which is valid for most beam-driven plasma wakefield accelerators), and noting that the cooperation length $l_\mathrm{c} \ll \lambda_\mathrm{p}$, such that $E_z$ varies approximately linearly over one cooperation length, the growth of the relative slice energy spread of the witness beam can be written as \cite{liu2025improvement}
\begin{equation}
    \delta_{\gamma_\mathrm{b}} \equiv \frac{\sigma_{\gamma_\mathrm{b}}}{\gamma_\mathrm{b}} = \frac{\sqrt{ \left(\sigma_{\gamma_{\mathrm{b}0}} m c^2\right)^2+\left(e\frac{l_\mathrm{c}}{2\sqrt{3}} \frac{\mathrm{d}E_z}{\mathrm{d}\xi}L\right)^2}}{ \gamma_\mathrm{b0}mc^2 + eE_z L},
    \label{energy_spread}
\end{equation}
where $\gamma_{\mathrm{b}0}mc^2$ and $\sigma_{\gamma_{\mathrm{b}0}}mc^2$ are the initial average energy and rms energy spread of the witness beam, $e$ is the electron charge, $L$ is the acceleration distance, and a flattop slice with a total thickness of $l_\mathrm{c}$ is assumed. Because the cooperation length depends on both the electron beam and undulator parameters, its precise value is not uniquely fixed. Guided by the discussion in Sec. IV, we adopt $l_\mathrm{c}\approx60~\nano\meter$ and $\rho\approx1.6\times10^{-4}$ as a representative value for the following calculations. Without the witness beam, $E_z$ falls approximately linearly with $\frac{\mathrm{d}E_z}{d\xi}\sim\mathrm{GeV/m/\micro m}$ in the acceleration region [red line in Fig. \ref{fig:wakefield_and_parameter_scan}(a)]. This would significantly increase the slice energy spread after a meter-scale plasma. A witness beam with a properly chosen delay and peak current can absorb the wake energy and flatten the $E_z$ profile to reduce $\left| \frac{\mathrm{d}E_z}{\mathrm{d}\xi} \right|$ and the corresponding $\delta_W$ after acceleration.

We introduce $\eta$ to quantify the fraction of electrons that satisfy the energy spread requirement after acceleration and perform one-step QuickPIC simulations to study the TR and $\eta$ using Eq. \ref{energy_spread}. As shown in Figs. \ref{fig:wakefield_and_parameter_scan}(b)-(e), the transformer ratio exhibits a clear dependence on the witness beam current and beam separation: lower current or larger separation results in a higher TR. The achieved TR ranges from $\sim2$ to $\sim4$ across the four cases, which is higher than that obtained with a driver of $\sigma_{z,\mathrm{d}}=16~\micro\meter$, for which $\text{TR}\approx1$ \cite{liu2025improvement}. TR can be further improved by using a shorter witness beam and/or a longer driver beam. These results are consistent with the beam loading analysis in the nonlinear plasma wake \cite{PhysRevLett.101.145002}. A larger $\Delta$ positions the witness beam closer to the wake tail, where $E_z$ is stronger; meanwhile, a smaller $I_\mathrm{w}$ or a shorter $\sigma_{z,\mathrm{w}}$ causes less modification of the unloaded high acceleration gradient. 

We divide the beam loading into three regimes \cite{liu2025improvement}: the underloaded regime where $\frac{\mathrm{d}E_z}{\mathrm{d}\xi}>0$ over the witness beam, the well-loaded regime where $\frac{\mathrm{d}E_z}{\mathrm{d}\xi}\approx 0$ and the overloaded regime with $\frac{\mathrm{d}E_z}{\mathrm{d}\xi}> 0$ close to the middle of the beam and away from the beam center. When $I_\mathrm{w}$ is low ($\lesssim 1.0~\mathrm{kA}$ for Case \Romannum{1} and \Romannum{2}, and $\lesssim 1.7~\mathrm{kA}$ for Case \Romannum{3} and \Romannum{4}), the witness beam carries insufficient charge to significantly modify the unloaded $E_z$ profile, leaving $\frac{\mathrm{d}E_z}{\mathrm{d}\xi}$ large and resulting in a slice energy spread that exceeds the Pierce parameter $\rho$ for most electrons [the deep blue regions in Figs. \ref{fig:wakefield_and_parameter_scan}(b)-(e)]. As $I_\mathrm{w}$ increases, $\frac{\mathrm{d}E_z}{\mathrm{d}\xi}$ decreases and $\eta$ increases correspondingly. When approaching the well-loaded regime, $\eta$ reaches a relatively high level and remains largely insensitive to small variations in $I_\mathrm{w}$ and $\Delta$. As can be seen from Figs. \ref{fig:wakefield_and_parameter_scan}(b)-(e), the current $I_\mathrm{w}$ achieving the maximum $\eta$ decreases as $\Delta$ increases \cite{PhysRevLett.101.145002, liu2025improvement}. A further increase of $I_\mathrm{w}$ results in the overloaded regime [upper right corners in Figs. \ref{fig:wakefield_and_parameter_scan}(b)-(e)] and $\eta$ decreases. 

Compared with Cases~\Romannum{1} and \Romannum{2}, the high-$\eta$ region in Case~\Romannum{3} and Case \Romannum{4} corresponds to a substantially larger $I_\mathrm{w}$, since the stronger driver excites a larger plasma bubble with a steeper field gradient at the tail, requiring a higher $I_\mathrm{w}$ to reach the well-loaded regime. Since the saturation power of FELs is $P_\mathrm{sat} \propto I_\mathrm{w}^{4/3}$, a large $I_\mathrm{w}$ is favorable for FEL performance \cite{pellegrini2016physics,HuangRevofFEL2007}. Point O in Case~\Romannum{3} achieves $\text{TR}\approx3.0$ and $\eta\approx 64\%$ at a peak current of $I_\mathrm{w}=2.2~\mathrm{kA}$, is well separated from the low-$\eta$ region, and is thus chosen for the subsequent full PBA simulation with QuickPIC. Throughout the PBA stage, the plasma wake structure remains nearly stationary, and the witness beam is continuously accelerated with an acceleration gradient of $15.9~\mathrm{GV/m}$ until the driver beam is depleted at a plasma length of $L=1.30~\mathrm{m}$. The center energy of the witness beam $W_\mathrm{c}$ reaches $ 28.9~\mathrm{GeV}$ at the exit of the PBA stage, which is $3.6$ times the initial energy. The current profile and the normalized emittance of the witness beam are preserved during the acceleration. Its relative energy spread closely matches the prediction obtained from Eq. \eqref{energy_spread} using the one-step simulation results, and the usable electron fraction $\eta \approx68\%$ is also in good agreement with the prediction of the one-step simulation.

\section{HI suppression using transversely asymmetric drivers}
 
The results above demonstrate that, under ideal driver--witness alignment and with axisymmetric beam distributions, a round driver can provide the required energy gain while preserving the witness-beam quality needed for subsequent XFEL operation. In practice, however, neither perfect alignment nor exact beam axisymmetry can be guaranteed. In the high-TR configuration studied here, finite initial driver--witness misalignment or asymmetric beam distributions can seed HI and rapidly degrade the witness-beam quality \cite{Whittum1991,Huang2007}. The HI of the driver can be mitigated by an initially large energy chirp or slice energy spread, as well as by effects that develop during PBA, including energy chirp, slice-energy-spread growth, and the nonlinear, axially varying focusing wakefield at the front of the driver \cite{PhysRevLett.118.174801, 2018IEEEmitigation, PhysRevAccelBeams.22.031302, Finnerud2025instability, PhysRevLett.119.244801}. These mechanisms, however, are unsuitable for the witness beam, whose projected energy spread, slice energy spread, and emittance must be preserved. Here, we utilize the axially varying focusing gradient of an elliptical nonlinear plasma wake driven by an elliptical driver \cite{2018IEEEmitigation,Manwani25PRL} to suppress the HI of the witness beam via Balakin–Novokhatsky–Smirnov (BNS) damping \cite{BNS1983}. We focus on HI seeded by a misalignment between the driver and the witness, and the same approach can be directly extended to suppress HI seeded by witness beam asymmetry. 

\begin{figure}[htbp]
    \centering    \includegraphics[width=1.0\columnwidth]{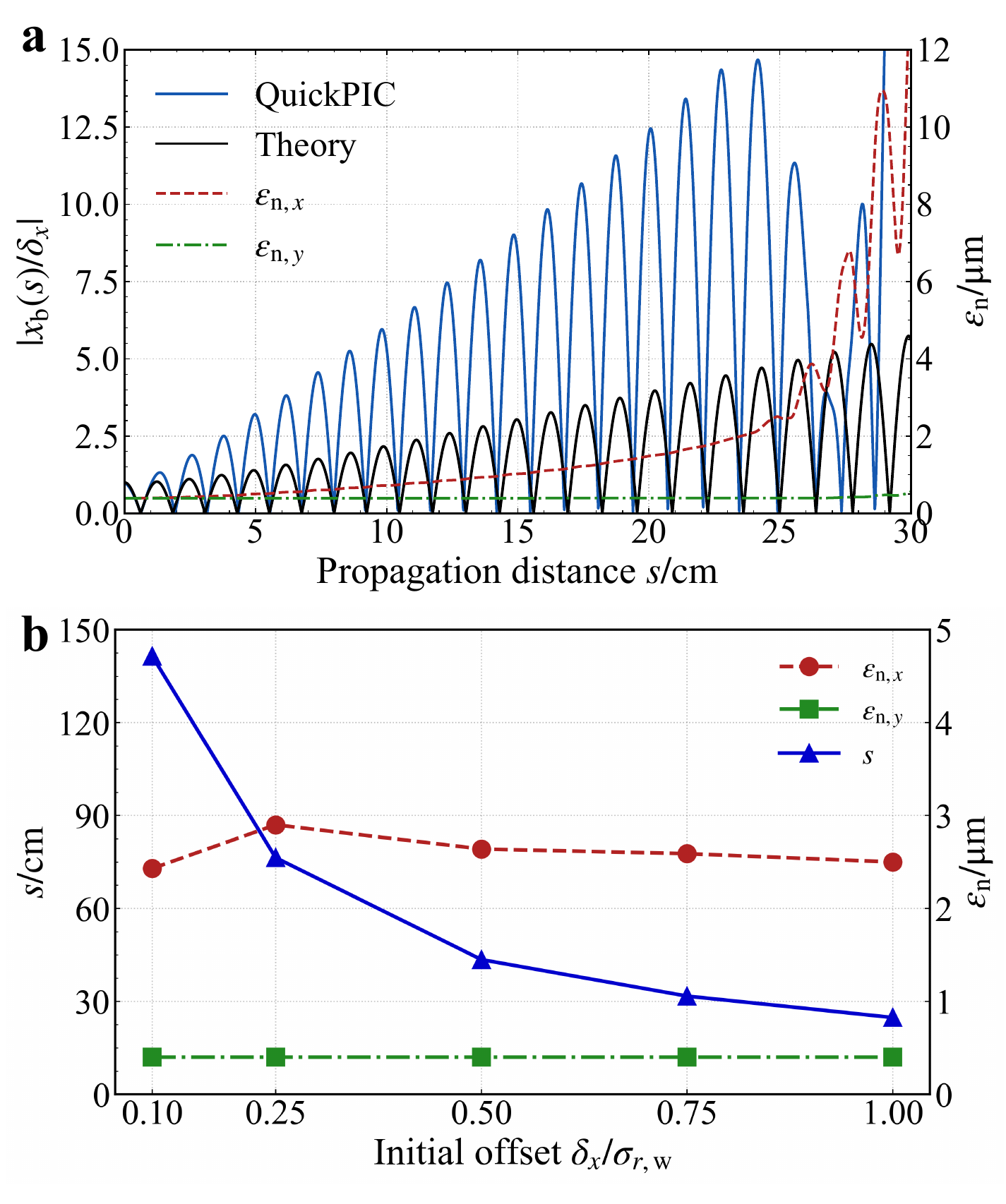}
    \caption{Witness-beam HI seeded by driver--witness misalignment. (a) Evolution of the slice centroid at $\xi=\xi_\mathrm{w}+\sigma_{z,\mathrm{w}}$ (blue solid line) and projected emittances within $\xi_\mathrm{w}\pm\sigma_{z,\mathrm{w}}$ (red dashed and green dash-dotted lines) for an initial offset $\delta_x=\sigma_{r,\mathrm{w}}$. The black solid line shows the centroid evolution calculated using the Huang model \cite{Huang2007}. (b) Dependence of the propagation distance at which the $\xi_\mathrm{w}+\sigma_{z,\mathrm{w}}$ slice reaches the ion column boundary on the initial offset (blue solid line), together with the projected emittances evaluated at that distance (red dashed and green dash-dotted lines). Immobile ions are used here. }
    \label{fig:HI}
\end{figure}

We introduce an initial transverse offset $\delta_x=\sigma_{r,\mathrm{w}}$ between the witness and driver at point O discussed in Sec. II, and use the slice at $\xi=\xi_\mathrm{w} + \sigma_{z,\mathrm{w}}$ to illustrate the development of HI. Because the witness beam is located near the rear of the nonlinear wake, where the ion channel radius is $r_\mathrm{b}\sim O(0.1)k_\mathrm{p}^{-1}$, it couples strongly to the wake boundary, leading to rapid growth of the HI \cite{Huang2007}. As shown by the blue line in Fig. \ref{fig:HI}(a), the oscillation amplitude of slice centroid $|x_\mathrm{b}|$ grows rapidly and reaches the boundary of the ion column at $s\approx24~\centi\meter$. Electrons then begin to be lost, causing the centroid oscillation amplitude to decrease. The HI also drives growth of the projected emittance in the oscillation direction: $\epsilon_{\mathrm{n},x}$ increases from its initial value of $0.4~\micro\meter$ to $2.5~\micro\meter$ at $s\approx24~\centi\meter$, whereas $\epsilon_{\mathrm{n},y}$ remains nearly unchanged. For comparison, the black solid line in Fig.~2(a) shows the centroid evolution predicted by the Huang model \cite{Huang2007}, using the channel radius and boundary pseudo-potential extracted from simulations and accounting for witness acceleration. The witness is located near the wake tail, where the wake radius varies rapidly along the longitudinal direction, thereby violating the short-pulse approximation adopted in the model. Consequently, the model underestimates the growth of the centroid oscillation amplitude.

We then scan the initial offset and record the propagation distance at which the selected slice reaches the ion column boundary, shown by the blue solid line in Fig. \ref{fig:HI}(b). Even for an offset of $\delta_x=0.1\sigma_{r,\mathrm{w}}$, the selected witness slice reaches the ion column boundary near the end of the plasma. For all values of $\delta_x$ considered, $\epsilon_{\mathrm{n},x}$ grows to approximately $2.5~\micro\meter$, while $\epsilon_{\mathrm{n},y}$ remains close to its initial value. These simulation results demonstrate that the HI can develop rapidly even from a small initial offset, causing electron loss and emittance growth that would preclude the use of the accelerated beam for XFEL applications.

\begin{figure}[!t]
    \centering
    \includegraphics[width=1.0\columnwidth]{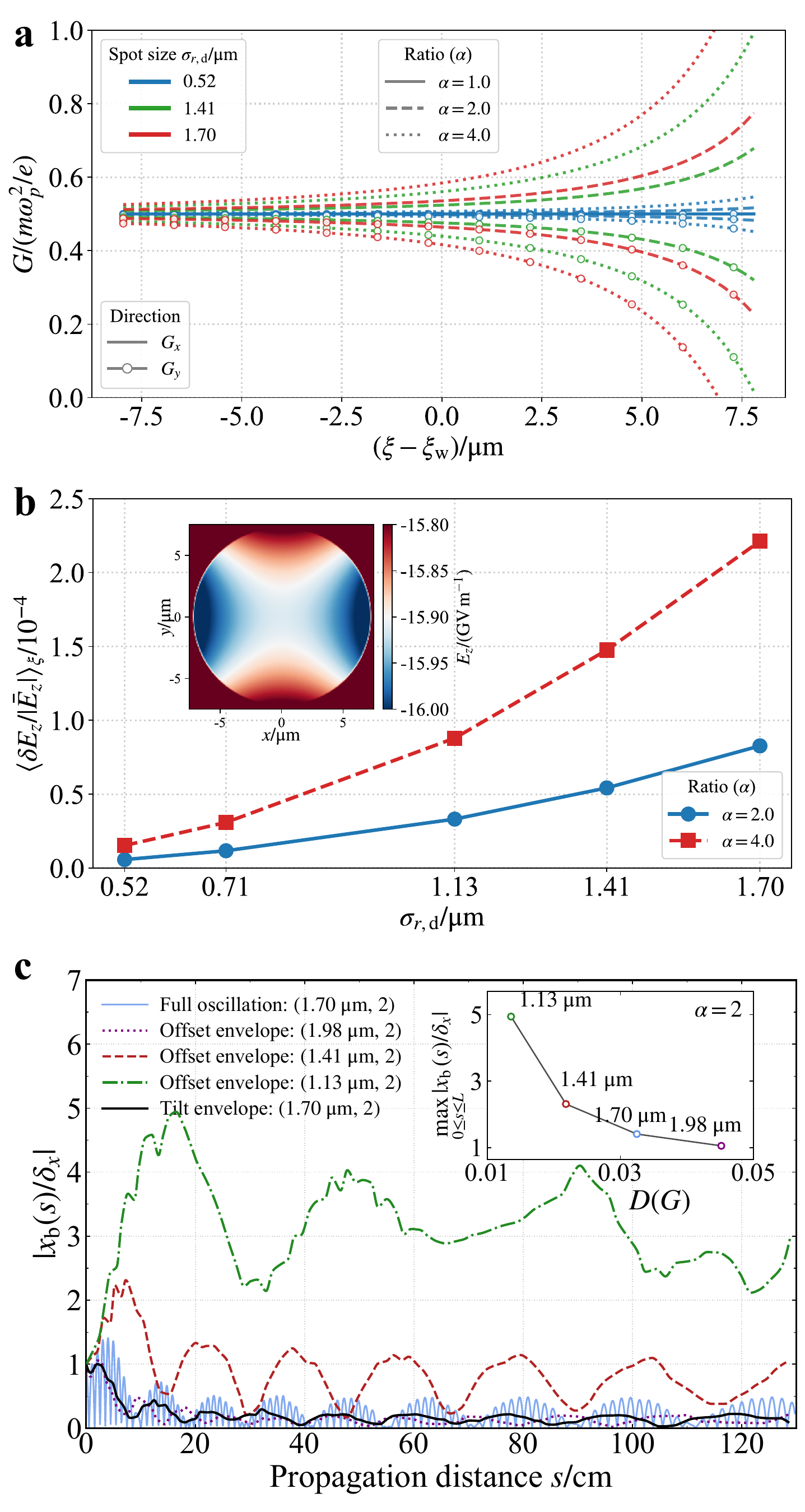}
    \caption{HI suppression by using transversely asymmetric drivers. (a) Axial variation of the focusing gradients for different driver configurations. (b) Relative transverse nonuniformity of the accelerating field for different driver configurations. The inset shows the transverse distribution of $E_z$ at $\xi=\xi_\mathrm{w}$ for $\sigma_{r,\mathrm{d}}=1.41~\micro\meter$ and $\alpha=2$. (c) Slice centroid evolution at $\xi=\xi_\mathrm{w}+\sigma_{z,\mathrm{w}}$ during the PBA stage for an initial offset $\delta_x=\sigma_{r,\mathrm{w}}$. The full centroid oscillation is shown for $\sigma_{r,\mathrm{d}}=1.70~\micro\meter$ and $\alpha=2$ (blue solid curve), whereas only the low-frequency envelopes are shown for the other driver configurations. The black solid curve shows the centroid envelope for an initially tilted witness when $(\sigma_{r,\mathrm{d}},\alpha)=(1.70~\micro\meter,2)$. The inset shows the maximum centroid amplification $A_{\max}$ versus the relative rms focusing-gradient variation $D(G)$ for different driver spot sizes, with $\alpha=2$; the point labels indicate $\sigma_{r,\mathrm{d}}$.} 
    \label{fig:asy_fields}
\end{figure}

The space charge field of an electron beam driver with an elliptical transverse profile is also elliptical. Take a Gaussian distribution $n_\mathrm{d}=n_\mathrm{d0}\exp\left(-\frac{x^2/\alpha + \alpha y^2}{2\sigma_{r,\mathrm{d}}^2} \right)$ as an example, $E_x\approx \frac{en_\mathrm{d0} \sigma_{r,\mathrm{d}}^2 x}{\epsilon_0 (x^2+y^2)}\left[ 1+ \frac{\sigma_{r,\mathrm{d}}^2 (\alpha - 1/\alpha )(x^2 -3y^2)}{(x^2 + y^2 )^2} \right], E_y \approx \frac{en_\mathrm{d0}\sigma_{r,\mathrm{d}}^2 y}{\epsilon_0 (x^2+y^2)}\left[ 1 + \frac{\sigma_{r,\mathrm{d}}^2 (\alpha -1/\alpha)(3x^2 -y^2)}{(x^2 + y^2 )^2} \right]$ when $x, y \gg \sigma_{r,\mathrm{d}}$, where $\alpha$ is aspect ratio of the driver. The second terms in the square brackets represent the nonaxisymmetric field components induced by the elliptical profile, which scale linearly with $\sigma_{r,\mathrm{d}}^2$ and $\alpha - \frac{1}{\alpha}$. Such a space charge field structure drives a blowout plasma wake with an elliptical boundary \cite{2018IEEEmitigation, Manwani25PRL}. Consequently, the gradient of the linear focusing wakefield varies along the axial direction, rather than remaining constant at $\frac{\mathrm{d}W_x}{\mathrm{d}x}=\frac{\mathrm{d}W_y}{\mathrm{d}y}=-\frac{1}{2}\frac{m\omega_\mathrm{p}^2}{e}$ as in a round wake. Figure \ref{fig:asy_fields}(a) shows the profiles of $G_x =- \frac{\mathrm{d}W_x}{\mathrm{d}x}$ and $G_y =-  \frac{\mathrm{d}W_y}{\mathrm{d}y}$ in the witness beam region under different driver spot sizes and aspect ratios. In these scans, the driver emittances in both transverse directions are adjusted to maintain transverse beam matching, while the driver current is kept fixed. A larger $\sigma_{r,\mathrm{d}}$ or a higher $\alpha$ leads to a more significant variation of $G_{x,y}$ along $\xi$. This axially varying $G_{x,y}$ changes the Betatron oscillation frequencies of different witness slices and drives them out of the resonance, thereby providing BNS damping of the HI \cite{BNS1983}.

Meanwhile, the Panofsky-Wenzel theorem ensures that an axially varying focusing field must be accompanied by a transversely non-uniform accelerating gradient \cite{Panofsky_Wenzel_1956}, i.e., $\frac{\partial \vec{W}_\perp}{\partial \xi} = \nabla_\perp E_z$. The inset in Fig. \ref{fig:asy_fields}(b) shows the $E_z$ distribution in the $x-y$ plane, where a clear transverse nonuniformity is observed. To quantify the transverse nonuniformity of $E_z$, we define $\delta E_z(\xi) \equiv \sqrt{\frac{\int \mathrm{d}x\mathrm{d}y n_\mathrm{w}(E_z-\bar{E}_z)^2}{\int \mathrm{d}x\mathrm{d}y n_\mathrm{w}}}$, where $\bar{E}_z(\xi) \equiv \frac{\int \mathrm{d}x\mathrm{d}y n_\mathrm{w} E_z}{\int \mathrm{d}x\mathrm{d}y n_\mathrm{w}}$ is the average acceleration gradient experienced by the witness beam. Figure \ref{fig:asy_fields}(b) shows the relative nonuniformity, averaged over the witness beam current, for different drivers. The trend is consistent with Fig. \ref{fig:asy_fields}(a), i.e., a larger $\sigma_{r,\mathrm{d}}$ or a higher $\alpha$ causes a larger $|\langle\frac{\delta E_z}{\bar{E}_z}\rangle_\xi|$, where $\langle f\rangle_\xi \equiv\frac{\int\mathrm{d}\xi I_\mathrm{w} f}{\int\mathrm{d}\xi I_\mathrm{w}}$. Because the nonuniformity of $E_z$ in an elliptical wake scales as $x^2-y^2$, employing a tightly focused witness beam is \mbox{beneficial} for achieving a small $|\langle\frac{\delta E_z}{\bar{E}_z}\rangle_\xi|$.

With an elliptical driver, the centroid evolution at the witness beam slice $\xi=\xi_\mathrm{w}+\sigma_{z,\mathrm{w}}$ for an initial offset of $\delta_x=\sigma_{r,\mathrm{w}}$ is shown in Fig.~\ref{fig:asy_fields}(c). When $\sigma_{r,\mathrm{d}}=1.70~\micro\meter$ and $\alpha=2$, the slice centroid oscillation (blue solid line) is suppressed effectively with a maximum $x_\mathrm{b} \approx 1.5\delta_x$. In addition to the high-frequency betatron oscillation, the centroid exhibits a slow oscillation with a decaying amplitude \cite{PhysRevLett.118.174801, PhysRevLett.119.244801}. Numerical calculations of the coupled HI equations \cite{Huang2007} reproduce this beat-wave behavior and show that it originates from resonance detuning induced by the axially varying focusing gradient. The beat-wave period can be estimated from the difference in betatron wavenumber between the beam head and a slice located at $\xi=\xi_\mathrm{w}+\sigma_{z,\mathrm{w}}$. This difference is $\Delta k_\beta = \sqrt{\frac{G(\xi)/({m\omega_\mathrm{p}^2}/{e})}{\gamma_\mathrm{b}}}k_\mathrm{p} - \sqrt{\frac{G_0/({m\omega_\mathrm{p}^2}/{e})}{\gamma_\mathrm{b}}}k_\mathrm{p} \approx 0.34 \frac{k_\mathrm{p}}{\sqrt{2\gamma_\mathrm{b}}}$, where $G_0=0.5\frac{m\omega_\mathrm{p}^2}{e}$ and $G(\xi) = 0.9\frac{m\omega_\mathrm{p}^2}{e}$ is used to reflect the increase of the focusing gradient during driver propagation (as discussed below). The betatron phase difference between these two slices therefore accumulates to $2\pi$ over a propagation distance $ \frac{2\pi}{\Delta k_\beta}\approx 18 \sqrt{2\gamma_\mathrm{b}} k_\mathrm{p}^{-1}$. This estimate is in reasonable agreement with the simulation result, shown by the blue line in Fig. \ref{fig:asy_fields}(c).

The envelope of the centroid oscillation for other drivers is also shown in Fig. \ref{fig:asy_fields}(c). A smaller $\sigma_{r,\mathrm{d}}$ results in less effective suppression of HI (red dashed line and green dash-dotted line), and vice versa (purple dotted line). This trend is consistent with the gradient profiles shown in Fig. \ref{fig:asy_fields}(a). To quantify the dependence of HI suppression on the longitudinal variation of the focusing gradient, we define $D(G)=\sqrt{\langle[G-\langle G\rangle_\xi]^2\rangle_\xi}/\langle G\rangle_\xi$, where $G(\xi)=G_x(\xi)$ and the current-weighted average is taken over $\xi_\mathrm{w}-2\sigma_{z,\mathrm{w}}\leq\xi\leq\xi_\mathrm{w}+\sigma_{z,\mathrm{w}}$. The corresponding centroid amplification is characterized by $A_{\max}=\max_{0\leq s\leq L}|x_\mathrm{b}(s)/\delta_x|$ at $\xi=\xi_\mathrm{w}+\sigma_{z,\mathrm{w}}$. As shown in the inset of Fig.~3(c), increasing $D(G)$ from approximately $0.014$ to $0.045$ reduces $A_{\max}$ from $4.94$ to $1.06$, consistent with stronger HI suppression through betatron-frequency detuning. As the head of the driver expands during propagation, it provides an increasingly nonaxisymmetric field component, and the axial variation of $G_{x,y}$ (as well as the transverse nonuniformity of $E_z$) increases correspondingly. Simulations show that for $(\sigma_{r,\mathrm{d}}, \alpha)=(1.70~\micro\meter, 4)$, $G_y$ at $\xi_\mathrm{w}+\sigma_{z,\mathrm{w}}$ becomes negative after a propagation distance of $s\approx 1 ~\centi\meter$, causing substantial electron loss. This imposes a constraint on the transverse profile of the driver.

For subsequent XFEL operation, we need to examine the witness beam at the PBA exit in terms of its offset amplification, slice emittance and slice energy spread growth. The envelope of $\left|\frac{x_\mathrm{b}}{\delta_x}\right|$ at the plasma exit ($s=1.3~\meter$) is approximately $3.3$, $1.0$, $0.3$, and $0.2$ for $(\sigma_{r,\mathrm{d}},\alpha)=(1.13~\micro\mathrm{m}, 2)$, $(1.41~\micro\mathrm{m}, 2)$, $(1.70~\micro\mathrm{m}, 2)$, and $(1.98~\micro\mathrm{m}, 2)$, respectively. Thus, the position offset of the witness beam is not amplified for $\sigma_{r,\mathrm{d}}\gtrsim 1.41~\mu\mathrm{m}$ and $\alpha = 2$. Since each slice of the witness beam is not transversely matched to the local focusing gradient, a finite slice energy spread can cause a growth of the slice emittance \cite{xu2016matching}. However, since the relative slice energy spread is very small, $\sim10^{-4}$, the resulting slice emittance growth is negligible over the considered 1.3-meter-long plasma \cite{xu2016matching}. The slice energy spread grows due to the transverse non-uniformity of $E_z$ in the elliptical wake. Accordingly, the usable fraction $\eta$ decreases to $62\%$ when $(\sigma_{r,\mathrm{d}},\alpha)=(1.13~\micro\mathrm{m},2)$, and decreases further with increasing $\sigma_{r,\mathrm{d}}$, i.e., $\eta\approx 50\%$, $41\%$ and $30\%$ for $(1.41~\micro\mathrm{m},2)$, $(1.70~\micro\mathrm{m},2)$, and $(1.98~\micro\mathrm{m},2)$, respectively. Considering the trade-off between the slice emittance and energy spread growth of the witness beam, and the HI suppression, we select a driver with $\sigma_{r,\mathrm{d}}=1.70~\micro\meter$ and $\alpha=2$ [blue solid line in Fig. \ref{fig:asy_fields}(c)]. Note that the peak current of the witness beam is slightly increased from $2.20$ to $2.35~\kilo\ampere$ to compensate the change of on-axis $E_z$ caused by the elliptical driver, increasing the usable fraction $\eta$ to approximately $50\%$. 

To test whether HI suppression remains effective for a different type of initial seed, we also simulate an initially tilted witness when $(\sigma_{r,\mathrm{d}},\alpha)=(1.70~\micro\meter,2)$. Its initial centroid follows $x_\mathrm{b}(0,\xi)=-\sigma_{r,\mathrm{w}}[\xi-(\xi_\mathrm{w}-3\sigma_{z,\mathrm{w}})]/(4\sigma_{z,\mathrm{w}})$, giving an offset magnitude of $\sigma_{r,\mathrm{w}}$ at $\xi=\xi_\mathrm{w}+\sigma_{z,\mathrm{w}}$. The corresponding centroid envelope, shown by the black solid curve in Fig.~3(c), exhibits negligible amplification. Thus, HI suppression also persists for an initially tilted witness beam. Mobile lithium ions are included in the simulations, and the associated ion motion is found to be weak, with a negligible effect on the HI. Additional simulations further show that, although hydrogen ions can effectively suppress the HI, this suppression is accompanied by substantial emittance growth \cite{IonMotion2017}. 

\section{XFELs with $\gtrsim50$ keV photon energies}

At the PBA exit, the witness beam is strongly focused with $\beta_{x,y} \sim 7~\mathrm{mm}$, making it unsuitable for direct injection into the undulator. To address this mismatch, a magnetic transport line is designed to increase the transverse $\beta$-functions of the witness beam, match them to the undulator focusing lattice, and remove the energy-depleted driver beam \cite{liu2025improvement}. The detailed design of this transport line is described in Appendix B.

Figure \ref{fig:transport_slice} shows the slice properties at the end of the transport line for a witness beam accelerated in an elliptical PBA with an initial driver--witness misalignment of $\sigma_{r,\mathrm{w}}$. The current profile (black solid line) and the slice emittances in the two transverse planes (red and green solid lines) are well preserved over most of the witness bunch. The relative slice energy spread remains at the $\sim 10^{-4}$ level near the bunch center, as indicated by the blue dash-dotted line. Owing to the axially varying focusing gradient in the PBA, the witness beam is not transversely matched to the plasma wake. As a result, the Courant--Snyder parameters of the witness beam vary from slice to slice at the PBA exit. Consequently, the slice $\beta$-functions exhibit axial oscillations after transport, as shown by the red and green dashed curves. Near the bunch center, where the slice energy spread is close to its minimum, both $\beta_x$ and $\beta_y$ are approximately $45~\mathrm{m}$.

\begin{figure}[htbp!]
    \centering
    \includegraphics[width=1.0\columnwidth]{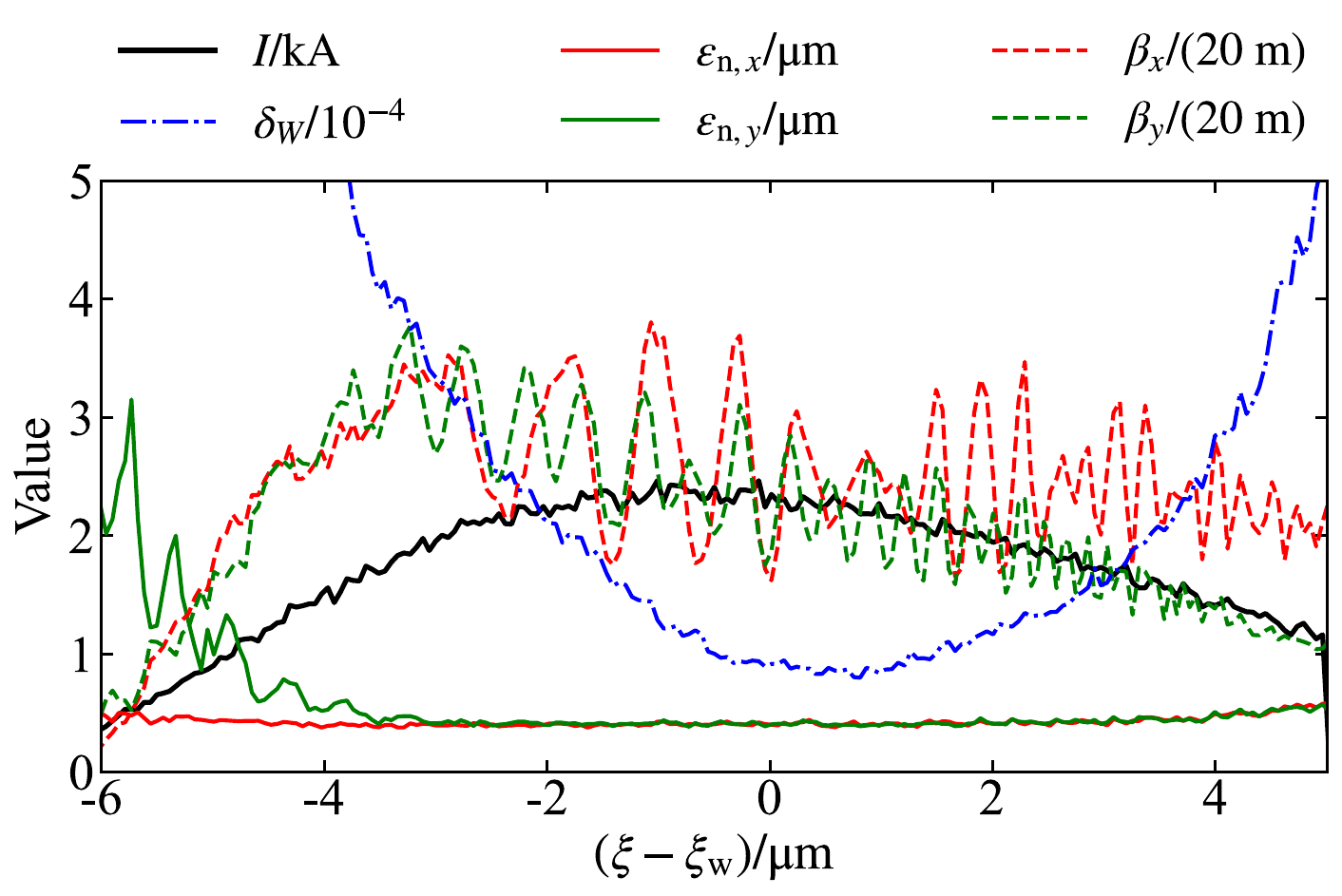}
    \caption{The slice current $I$, relative energy spread $\delta_W$, normalized emittance $\epsilon_\mathrm{n}$ and $\beta$-function of the witness bunch at the end of the beam transport line.}
    \label{fig:transport_slice}
\end{figure}

According to the undulator resonance condition, $\lambda_\mathrm{r} = \frac{\lambda_\mathrm{u}}{2\gamma_\mathrm{b}^2}\left( 1 + \frac{K^2}{2}\right)$, this $\sim 30$ GeV and high-quality electron beam can drive FEL lasing at high photon energies in an undulator. However, quantum diffusion induced by continuous spontaneous undulator radiation (SUR) reduces the electron average beam energy and increases its energy spread \cite{pellegrini2016physics,HuangRevofFEL2007}. While the energy loss can be compensated for by employing a tapered undulator \cite{variableFELpara}, the growth of the energy spread ultimately limits the achievable photon energy.

The growth of the energy spread induced by quantum diffusion is described as \cite{diffusion1996saldin} $\frac{\mathrm{d}  \sigma_{\gamma_\mathrm{b}}^2 }{\mathrm{d} z_\mathrm{u}} = \frac{7}{15}\lambda_\mathrm{c} r_e \gamma_\mathrm{b}^4 k_\mathrm{u}^3 K^2 F\left(\frac{K}{\sqrt{2}}\right)$, where $r_e$ is the classical electron radius, $\lambda_\mathrm{c} = \frac{\hbar}{mc}$ is the reduced Compton wavelength, and $F(x)=1.70x+\frac{1}{1+1.88x+0.8x^2}$ (for planar undulator) is a fit to simplify the dependence on various Bessel functions. As the FEL instability develops, quantum diffusion increases the beam energy spread; once the induced relative energy spread approaches the FEL Pierce parameter $\rho$ before saturation, the FEL performance is strongly degraded. In 1D FEL theory, $\rho = \left(\frac{e^2 K^2[\mathrm{JJ}]^2n_\mathrm{b}}{32\epsilon_0\gamma_\mathrm{b}^3m c^2 k_\mathrm{u}^2}\right)^{1/3}$, where $n_\mathrm{b}$ is the peak beam density, and $[\mathrm{JJ}]\sim 1$ is the coupling factor between the electrons and the radiation field in a planar undulator whose value depends on $K$ \cite{pellegrini2016physics}. For a fixed target radiation wavelength, for which $\lambda_\mathrm{u}\propto (1+K^2/2)^{-1}$, numerical calculations indicate that a larger $\lambda_\mathrm{u}$ and a smaller $K$ are usually favorable for FEL gain.

\begin{figure}[htbp!]
    \centering
    \includegraphics[width=1.0\columnwidth]{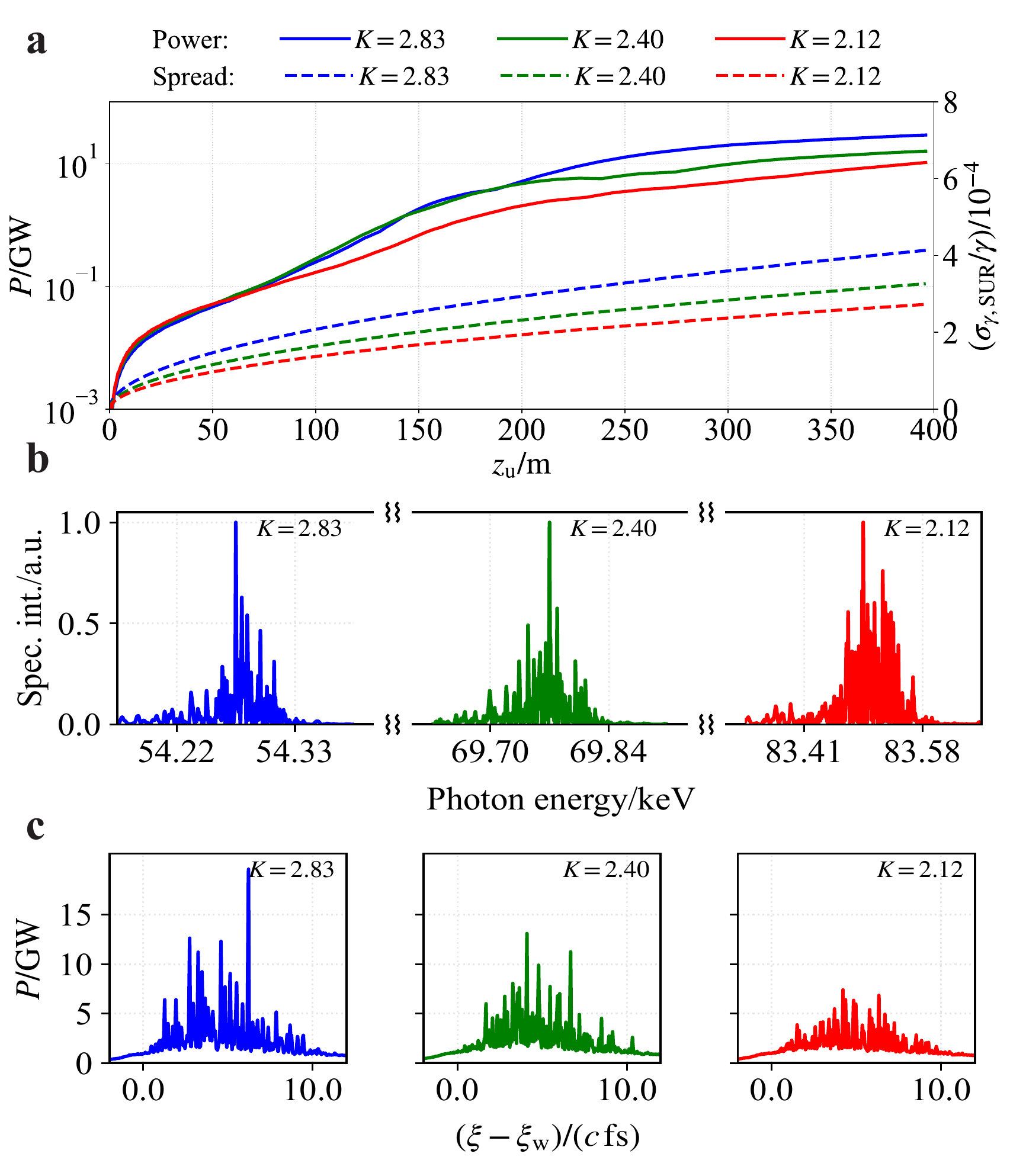}
    \caption{High-photon-energy XFEL performance. (a) Radiation peak power (solid curves) and relative energy spread induced by quantum diffusion (dashed curves) along the undulator for $K=2.83$, $2.40$, and $2.12$, with the undulator period fixed at $\lambda_{\mathrm{u}}=0.03~\mathrm{m}$. (b) Corresponding normalized radiation spectra evaluated near saturation, at $z_{\mathrm{u}}=300~\mathrm{m}$ for $K=2.83$ and $z_{\mathrm{u}}=350~\mathrm{m}$ for $K=2.40$ and $2.12$. (c) Corresponding temporal power profiles evaluated at the same undulator positions as in (b).}
    \label{fig:FEL_results}
\end{figure}

To assess the XFEL performance driven by the accelerated witness beam, we perform 3D FEL simulations with GENESIS for high-photon-energy XFEL operation, including quantum diffusion. For simplicity, we choose an undulator period of $\lambda_\mathrm{u}=3~\centi\meter$ and vary the undulator deflection parameter $K$ to vary the radiation wavelength. A stepwise linear taper is applied from the undulator entrance, with $K$ decreasing by $1.32\times10^{-4}$, $8.75\times10^{-5}$, and $6.45\times10^{-5}$ per $1.98~\meter$ module for the initial $K=2.83$, $2.40$, and $2.12$ cases, respectively, thereby keeping $(1+K^2/2)/\gamma_\mathrm{b}^2$ approximately constant and maintaining the FEL resonance condition. The solid lines in Fig. \ref{fig:FEL_results}(a) show the growth of the peak radiation power for three representative cases whereas the dashed lines indicate the quantum-diffusion-induced energy spread growth. 

For $\lambda_\mathrm{u}=3~\centi\meter$ and $K=2.83$, the resonant photon energy is $E_\mathrm{r}=54.3~\kilo\electronvolt$. The 3D Pierce parameter estimated using Ming Xie's fitting formula \cite{MingXie} is $\rho \simeq 1.6\times 10^{-4}$. Fitting the simulated power growth with $P(z_\mathrm{u})\propto\exp(z_\mathrm{u}/L_\mathrm{g})$ over the interval $100~\mathrm{m}\leq z_\mathrm{u}\leq140~\mathrm{m}$ gives a power gain length $L_\mathrm{g}$ corresponding to an effective Pierce parameter $\rho_\mathrm{eff}\equiv \lambda_\mathrm{u}/(4\pi\sqrt{3}L_\mathrm{g})\approx0.54\times10^{-4}$, approximately a factor of three smaller than the Ming Xie estimate. This reduction is mainly caused by the SUR-induced beam energy spread growth, as shown by the blue dashed line in Fig. \ref{fig:FEL_results}(a). The radiation power saturates at approximately $z_\mathrm{u}=300~\meter$, reaching a peak value of $P\approx 19.6~\giga\watt$.

As $K$ is decreased to 2.40 and 2.12, the corresponding photon energy increases to 69.7 and 83.5 keV and the Pierce parameter given by the Ming Xie's formula decreases to $1.3\times 10^{-4}$ and $1.1\times 10^{-4}$. As indicated by the dashed lines in Fig. \ref{fig:FEL_results}(a), the energy spread growth is also slowed compared with the $K=2.83$ case. In both cases, the peak power saturates at around $z_\mathrm{u}=350~\meter$, reaching $P \approx 13.1$ and $7.4~\giga\watt$, respectively.

The radiation spectra at saturation are shown in Fig. \ref{fig:FEL_results}(b). They exhibit pronounced multi-spike structures, with relative rms bandwidths $\Delta\omega/\langle\omega\rangle$ of $7.3\times10^{-4}$, $7.2\times10^{-4}$, and $7.5\times10^{-4}$ for $K=2.83$, $2.40$, and $2.12$, respectively. These bandwidths are approximately $13.6$, $13.9$, and $20.3$ times the corresponding effective Pierce parameters inferred from fits to the simulated exponential power growth. The larger full-pulse bandwidths arise from the residual longitudinal energy chirp acquired during the PBA stage: different slices have different resonant frequencies, introducing additional spectral broadening beyond the intrinsic SASE bandwidth \cite{HuangRevofFEL2007,liu2025improvement}. The temporal pulse profiles for these three cases are shown in Fig. \ref{fig:FEL_results}(c). The radiation pulse develops around the beam center where the current is high and the energy spread is small. In all three cases, the pulse duration is approximately $10~\mathrm{fs}$. At the corresponding saturation distances, the pulse energies are estimated to be $33.0$, $32.2$, and $26.2~\micro\mathrm{J}$ for photon energies of $54.3$, $69.7$, and $83.5~\mathrm{keV}$, respectively. These pulse energies correspond to approximately $3.8\times10^{9}$, $2.9\times10^{9}$, and $2.0\times10^{9}$ photons per pulse, respectively.

\section{Discussion}
A high transformer ratio requires a long driver beam and a short witness beam, with the driver carrying substantially more charge than the witness. Overall, only approximately $5\%$ of the charge delivered by the RF accelerator contributes to FEL lasing in the undulator and generates x-rays with high photon energies. In this sense, the PBA acts as an effective photon energy up-conversion stage: it sacrifices the number of radiating electrons in exchange for a substantial increase in photon energy. Without the afterburner, all electrons from the RF accelerator radiate in the undulator, but at relatively low photon energies. With the afterburner, only $\sim 5\%$ of the electrons radiate, yet the emitted photons reach much higher energies. A crude estimate suggests that this trade-off is equivalent to converting approximately $10^3$ photons at several keV into one photon above $50~\kilo\electronvolt$.

In conclusion, we use 3D PIC simulations to demonstrate that 8 GeV electron beams from RF accelerators can be boosted to $\sim30$ GeV in a meter-long PBA with a transformer ratio of $\approx 3$. The beam quality can be preserved well even in the presence of alignment errors, with nearly half of the electrons satisfying the requirements of subsequent high-photon-energy XFELs. The misalignment between the driver and witness beams can trigger the HI and we employ an elliptical beam driver to drive an elliptical plasma wake which can suppress the HI of the witness beam effectively and maintain the beam quality. The parameter study presented in Sec. III shows that the driver spot size needs to be chosen properly to achieve both effective HI suppression and minimal growth of slice energy spread. This poses challenges for the electron beams delivered by the RF accelerators, i.e., the spot sizes of the driver are different along the two transverse directions, and the average spot size is approximately 6 times that of the witness beam. Our simulations show that 10 GW-class x-rays with $\gtrsim 50$ keV photon energy can be produced in undulators hundreds of meters long after accounting for quantum diffusion.

The TR of approximately 3 considered here represents an operating point compatible with the beam parameters and XFEL application targeted in this study. The maximum TR at which HI can still be effectively suppressed depends on the competition between HI growth and longitudinal betatron-frequency detuning, subject to the beam-quality constraints associated with the specific driver–witness configuration. Although longitudinally shaped beams can provide higher TRs \cite{Su2023PoP}, the effectiveness of the proposed suppression mechanism in such configurations requires separate assessment. For the high-photon-energy XFEL application considered here, maximizing the TR is not the primary design objective. A TR of approximately 3 already provides sufficient witness energy for the demonstrated XFEL operation above 50 keV, provided that the required beam quality is preserved. Increasing the TR and witness energy does not necessarily improve the usable XFEL output: for fixed undulator parameters and propagation length, quantum-diffusion-induced relative energy spread increases with electron energy, while the Pierce parameter decreases when the beam current and transverse beam size are held fixed. Quantum diffusion already noticeably affects FEL performance in the cases studied here. Improving beam brightness offers a promising route by increasing the Pierce parameter and shortening the saturation length, thereby reducing the accumulated quantum-diffusion-induced energy spread \cite{tan2025arXiv}.

\begin{acknowledgments}

This work was supported by the Fundamental and Interdisciplinary Disciplines Breakthrough Plan of the Ministry of Education of China-JYB2025XDXM204, the National Grand Instrument Project No. SQ2019YFF01014400, National Natural Science Foundation of China No. 12375147, Beijing outstanding young scientist project, and the Fundamental Research Funds for the Central Universities, Peking University. The simulations were supported by the High-performance Computing Platform of Peking University.

\end{acknowledgments}

\section*{Author contributions}

\subsection*{Conflict of Interest}
The authors have no conflicts to disclose.

\subsection*{Author Contributions}
\textbf{Songtang Han}: Conceptualization (equal); Data curation (lead); Formal analysis (lead); Investigation (lead); Methodology (equal); Validation (lead); Visualization (lead); Writing -- original draft (lead); Writing -- review \& editing (equal). 
\textbf{Letian Liu}: Formal analysis (equal); Investigation (equal); Methodology (equal); Software (supporting);  Writing -- review \& editing (supporting). 
\textbf{Zhiyan Yang}: Investigation (equal); Software (supporting); Writing -- review \& editing (supporting). 
\textbf{Zhuo Pan}: Investigation (equal); Methodology (equal); Writing -- review \& editing (supporting). 
\textbf{Xueqing Yan}: Funding acquisition (equal); Supervision (equal); Writing -- review \& editing (equal). 
\textbf{Xinlu Xu}: Conceptualization (lead); Formal analysis (equal); Funding acquisition (equal); Investigation (equal); Methodology (equal); Resources (lead); Supervision (lead); Writing -- original draft (equal); Writing -- review \& editing (equal).

\section*{Data availability}
The data that support the findings of this study are available from the corresponding author upon reasonable request.

\appendix
\section{Setup of the simulations}
All simulations of the PBA stage are performed with the quasi-static 3D PIC code QuickPIC \cite{QuickPIC2006,an2013improved}. A moving window propagating at the speed of light in vacuum $c$ is adopted, with a box size of $160~\micro\meter\times160~\micro\meter\times260~\micro\meter$, and is discretized into $1024\times1024\times1024$ cells in the $x$, $y$, and $\xi$ directions, respectively. A convergence test on the grid resolution is performed to verify that the chosen grid size is appropriate. Both the driver and witness beams are represented by approximately $4.3\times10^6$ macro particles. The plasma electrons and ions are initialized with 4 macro particles per cell. In the quasi-static algorithm, as the driver-witness bunch pair passes through each plasma slice, the plasma response is solved and the resulting electromagnetic fields are then used to advance the positions and momenta of the beam particles. In the simulations presented here, the bunch pair is advanced with a longitudinal step of $2~\mathrm{mm}$.

The beam transport is simulated using the particle tracking code ELEGANT \cite{borland2000elegant} as detailed in Appendix B. The FEL process is simulated with the 3D time-dependent FEL code GENESIS~\cite{reiche1999genesis}. The six-dimensional phase-space distribution of the transported witness beam is imported into GENESIS at the undulator entrance. To preserve the transverse beam size along the undulator, a FODO focusing lattice is included in the GENESIS simulations. Guided by the optimum $\beta$-functions obtained from the Ming Xie fitting formula, the focusing and defocusing quadrupoles in the FODO lattice are chosen to have a length of $4~\mathrm{cm}$ and a gradient of $106~\mathrm{T/m}$, with each quadrupole pair separated by a $1.98~\mathrm{m}$-long undulator module. The GENESIS simulations include both the average energy loss and the energy spread growth induced by SUR, with a stepwise linear undulator taper applied to compensate for the average energy loss, as described in Section IV.

\section{Beam transport from the PBA to the undulator}

\begin{figure}[htbp!]
    \centering
    \includegraphics[width=1.0\columnwidth]{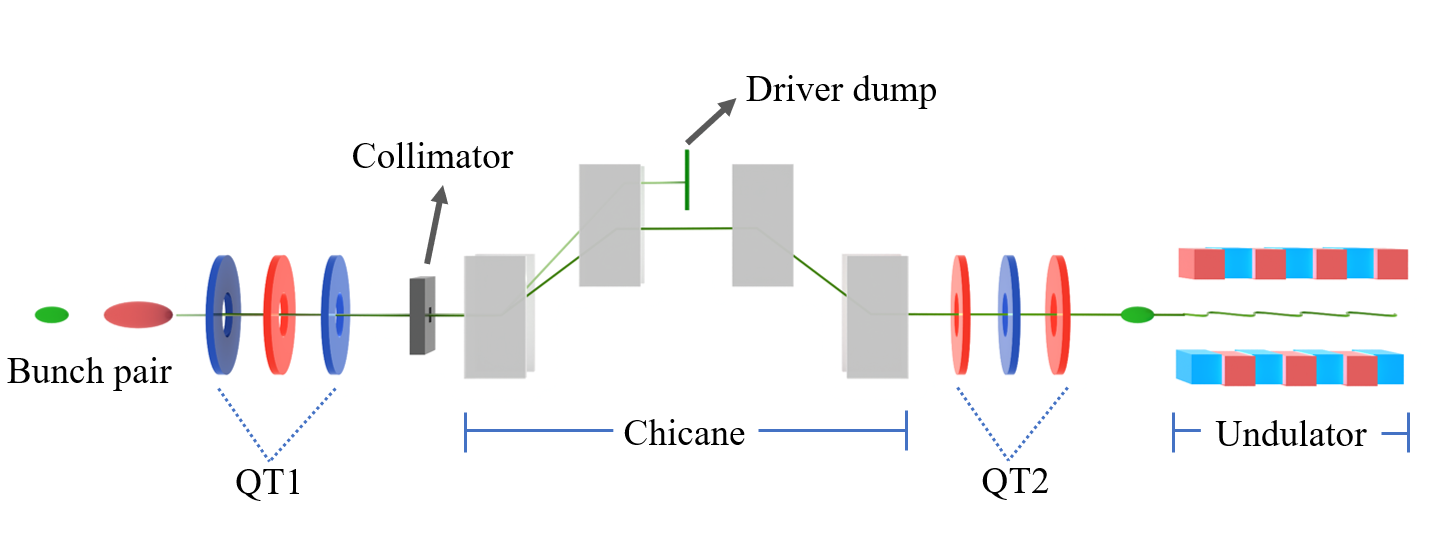}
    \caption{Schematic of the beam transport line from the plasma exit to the undulator entrance (not to scale). After the PBA stage, the accelerated witness beam and the energy-depleted driver beam are captured by the first quadrupole triplet, QT1. A rectangular collimator removes most of the driver electrons, while the residual driver is further separated from the witness beam in a symmetric C-chicane and dumped at the chicane center. The second quadrupole triplet, QT2, matches the witness beam to the target Twiss parameters at the undulator entrance for subsequent high-photon-energy FEL.}
    \label{fig:transport_line}
\end{figure}

A magnetic transport line is designed using the particle tracking code ELEGANT \cite{borland2000elegant}. As shown in Fig. \ref{fig:transport_line}, it consists of two quadrupole triplets (QT1 and QT2) and a symmetric C-chicane \cite{liu2025improvement}. QT1, in a D-F-D configuration, is placed $18~\mathrm{cm}$ downstream of the plasma exit, where `D/F' indicates defocusing/focusing in the $x$-direction. It consists of two defocusing quadrupoles (length $17.1~\mathrm{cm}$, gradient $-748.6~\mathrm{T/m}$) and one focusing quadrupole (length $34.2~\mathrm{cm}$, gradient $606.9~\mathrm{T/m}$), separated by $41~\mathrm{cm}$ drifts. The energy-depleted driver diverges more rapidly than the witness beam in QT1. A rectangular collimator of length $50~\mathrm{cm}$ with an aperture of $0.15~\mathrm{mm}\times0.6~\mathrm{mm}$ is placed $60~\mathrm{cm}$ downstream of QT1 to remove the majority of the driver electrons. The subsequent symmetric C-chicane consists of four dipoles, each with a length of $10~\mathrm{cm}$ and a magnetic field of 2.9 T, corresponding to a bending angle of $3~\mathrm{mrad}$ for 29.3 GeV electrons. The drift lengths between consecutive dipoles are $1.00~\mathrm{m}$, $0.50~\mathrm{m}$, and $1.00~\mathrm{m}$. A beam dump is placed at the center of the chicane to remove the residual driver, which is horizontally separated from the witness beam by the dispersion introduced in the first half of the chicane. QT2, in an F-D-F configuration, is placed $89~\mathrm{cm}$ downstream of the chicane exit. It consists of two focusing quadrupoles (length $13.8~\mathrm{cm}$, gradient $440.4~\mathrm{T/m}$) and one defocusing quadrupole (length $23.2~\mathrm{cm}$, gradient $-373.4~\mathrm{T/m}$), separated by $1.34~\mathrm{m}$ drifts. QT2 matches the witness beam to the target Twiss parameters $\beta_x = 48~\mathrm{m}$, $\beta_y = 44~\mathrm{m}$, $\alpha_x = 1.04$, and $\alpha_y = -0.96$ at the undulator entrance. These parameters are optimized using the Ming Xie's fitting formula \cite{MingXie} to maximize the 3D Pierce parameter $\rho$ and are approximately matched with the FODO along the undulator. Additional tracking simulations show that the transport largely preserves the entrance slice emittances under energy and emittance jitters, with moderate changes in the $\beta$-functions.

\bibliography{myref}

\end{document}